\documentclass[aps,prl,reprint,twocolumns,showpacs,superscriptaddress]{revtex4-1}
\usepackage{graphicx, amsmath}
\usepackage{amsfonts} % for mathbb
\usepackage{amssymb} % for intercal
\begin{document}

%-------------------------------------------------------------------------------
% Shortcut Commands
%-------------------------------------------------------------------------------

\newcommand{\comb}[2]{{\begin{pmatrix} #1 \\ #2 \end{pmatrix}}}
\newcommand{\braket}[2]{{\left\langle #1 \middle| #2 \right\rangle}}
\newcommand{\bra}[1]{{\left\langle #1 \right|}}
\newcommand{\ket}[1]{{\left| #1 \right\rangle}}
\newcommand{\ketbra}[2]{{\left| #1 \middle\rangle \middle \langle #2 \right|}}

%-------------------------------------------------------------------------------
% Front Matter
%-------------------------------------------------------------------------------

\title{Global Symmetry is Unnecessary for Fast Quantum Search}

\author{Jonatan Janmark}
	\email{jjanmark@kth.se}
	\affiliation{Department of Physics, KTH Royal Institute of Technology, 106 91 Stockholm, Sweden}
	\affiliation{Department of Mathematics, University of California, San Diego, La Jolla, CA 92093-0112}
\author{David A. Meyer}
	\email{dmeyer@math.ucsd.edu}
	\affiliation{Department of Mathematics, University of California, San Diego, La Jolla, CA 92093-0112}
\author{Thomas G. Wong}
	\email{tgw002@physics.ucsd.edu}
	\affiliation{Department of Physics, University of California, San Diego, La Jolla, CA 92093-0354}

\begin{abstract}
	Grover's quantum search algorithm can be formulated as a quantum particle randomly walking on the (highly symmetric) complete graph, with one vertex marked by a nonzero potential. From an initial equal superposition, the state evolves in a two-dimensional subspace. Strongly regular graphs have a local symmetry that ensures that the state evolves in a \emph{three}-dimensional subspace, but most have no \emph{global} symmetry. Using degenerate perturbation theory, we show that quantum random walk search on known families of strongly regular graphs nevertheless achieves the full quantum speedup of $\Theta(\sqrt{N})$, disproving the intuition that fast quantum search requires global symmetry.
\end{abstract}

\pacs{03.67.Ac, 02.10.Ox}

\maketitle

%-------------------------------------------------------------------------------
% Main Matter
%-------------------------------------------------------------------------------

\section{Introduction}

\begin{figure}
\begin{center}
	\includegraphics[width=1in]{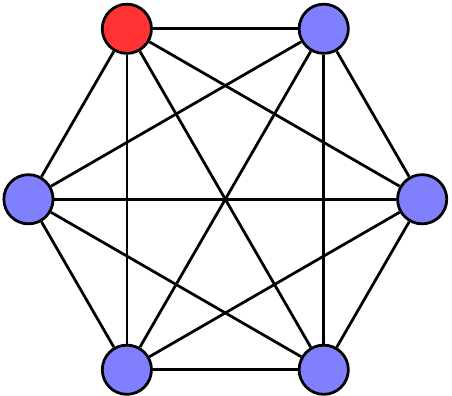}
	\includegraphics[width=1in]{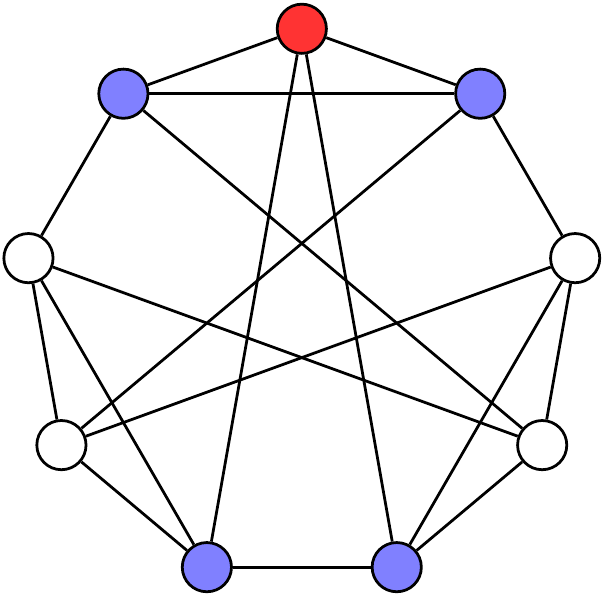}
	\includegraphics[width=1in]{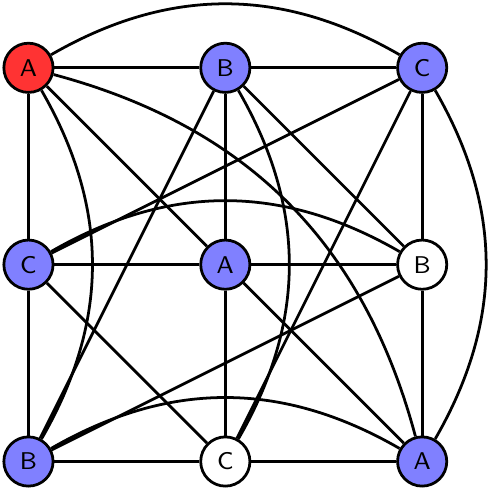}
	\caption{\label{fig:graphs}From left to right: the complete graph with $6$ vertices, the Paley graph with parameters (9,4,1,2), and the Latin square graph with parameters (9,6,3,6). Without loss of generality, a ``marked'' vertex is colored red, vertices adjacent to it are colored blue, and vertices not adjacent to it are colored white.}
\end{center}
\end{figure}

While Grover's algorithm was originally proposed as a digital, or discrete-time, algorithm \cite{Grover1996}, Farhi and Gutmann formulated it as an equivalent analog, or continuous-time, algorithm \cite{FG1998}. We use Childs and Goldstone's notation and interpretation of this algorithm \cite{CG2004} as a quantum randomly walking particle on the complete graph of $N$ vertices, an example of which is shown in Fig.~\ref{fig:graphs}.

The $N$ vertices of the graph label computational basis states $\{ \ket{0}, \dots, \ket{N-1} \}$ of an $N$-dimensional Hilbert space. The initial state $\ket{\psi(0)}$ is an equal superposition $\ket{s}$ of all these basis states:
\[ \ket{\psi(0)} = \ket{s} = \frac{1}{\sqrt{N}} \sum_{i=0}^{N-1} \ket{i}. \]
The goal is to find a particular ``marked'' basis state, which we label $\ket{w}$ and depict by a red vertex in Fig.~\ref{fig:graphs}. We search by evolving Schr\"odinger's equation with Hamiltonian
\begin{equation}
	H = -\gamma L - \ketbra{w}{w},
	\label{eq:QRW}
\end{equation}
where $\gamma$ is the amplitude per unit time of the randomly walking quantum particle transitioning from one vertex to another, $L$ is the graph Laplacian that effects a quantum random walk on the graph, and $\ketbra{w}{w}$ is a potential well at the marked vertex, which causes amplitude to accumulate there. More specifically, $L = A - D$, where $A_{ij} = 1$ if $(i,j) \in \mathcal{E}$, the set of edges of the graph, (and $0$ otherwise) is the adjacency matrix indicating which vertices are connected to one another, and $D_{ii} = \text{deg}(i)$ (and $0$ otherwise) is the degree matrix indicating how many neighbors each vertex has. Adding $N$ times the identity matrix, which is an unobservable rezeroing of energy or overall phase, yields $H = -\gamma N \ketbra{s}{s} - \ketbra{w}{w}$.

One might (correctly) reason that the success of the algorithm depends on the value of $\gamma$. When $\gamma$ takes its critical value of $\gamma_c = 1/N$, then $H = - \ketbra{s}{s} - \ketbra{w}{w}$, and its eigenstates are proportional to $\ket{s} \pm \ket{w}$ with corresponding eigenvalues $-1 \mp 1/\sqrt{N}$. So the Schr\"odinger evolution rotates the state from $\ket{s}$ to $\ket{w}$ in time $\pi / \Delta E = \pi \sqrt{N} / 2$, which is optimal \cite{Zalka1999}.

Degenerate perturbation theory \cite{Griffiths2005} provides an alternate method for analyzing this Hamiltonian. Since the non-marked vertices, depicted by the blue vertices in Fig.~\ref{fig:graphs}, evolve identically by symmetry, we can group them together:
\[ \ket{r} = \frac{1}{\sqrt{N-1}} \sum_{i \ne w} \ket{i}. \]
Then the system evolves in a two-dimensional subspace spanned by $\{\ket{w}, \ket{r}\}$. In this basis, the Hamiltonian is
\[ H = \begin{pmatrix}
	-(\gamma + 1) & -\gamma \sqrt{N-1} \\
	-\gamma \sqrt{N-1} & -\gamma(N-1) \\
\end{pmatrix}. \]
Assuming $N$ is large so that $N-1 \approx N$, we separate the Hamiltonian into leading and higher order terms of $O(1)$, $O(1/\sqrt{N})$, and $O(1/N)$:
\[ H = \underbrace{\begin{pmatrix}
	-1 & 0 \\
	0 & -\gamma N \\
\end{pmatrix}}_{H^{(0)}} + \underbrace{\begin{pmatrix}
	0 & -\gamma \sqrt{N} \\
	-\gamma \sqrt{N} & 0 \\
\end{pmatrix}}_{H^{(1)}} + \underbrace{\begin{pmatrix}
	-\gamma & 0 \\
	0 & 0 \\
\end{pmatrix}}_{H^{(2)}}. \]
In lowest order, the eigenstates of $H^{(0)}$ are $\ket{w}$ and $\ket{r}$ with corresponding eigenvalues $-1$ and $-\gamma N$. If the eigenvalues are nondegenerate, then since the initial superposition state $\ket{s}$ is approximately $\ket{r}$ for large $N$, the system will stay near its inital state, never having large projection on $\ket{w}$. If the eigenstates are degenerate (\textit{i.e.}, when $\gamma = \gamma_c = 1/N$), however, then the perturbation will cause the eigenstates of the perturbed system to be a superposition of $\ket{r}$ and $\ket{w}$:
\[ \ket{\psi_\pm} = \alpha_w \ket{w} + \alpha_r \ket{r}, \]
where the coefficients $\alpha_{w,r}$ and eigenvectors $E_\pm$ can be found by solving the eigenvalue problem
\[ \begin{pmatrix} H_{ww} & H_{wr} \\ H_{rw} & H_{rr} \end{pmatrix} \begin{pmatrix} \alpha_w \\ \alpha_r \end{pmatrix} = E_\pm \begin{pmatrix} \alpha_w \\ \alpha_r \end{pmatrix}, \]
where $H_{wr} = \langle w | H^{(0)} + H^{(1)} | r \rangle$, \textit{etc}. Solving this yields the eigenstates of the perturbed system: $\ket{\psi_\pm} = \frac{1}{\sqrt{2}} \left( \ket{w} \mp \ket{r} \right)$ with corresponding eigenvalues $E_\pm = -1 \pm 1/\sqrt{N}$.
Since $\ket{r} \approx \ket{s}$, the system evolves from $\ket{s}$ to $\ket{w}$ in time $t_* = \pi / \Delta E = \pi \sqrt{N} / 2$.

From the novel degenerate perturbation-theoretic perspective introduced above, the next step in difficulty would be search on a graph for which the state evolves in a three-dimensional subspace, for example that spanned by the marked vertex, the superposition of vertices adjacent to the marked vertex, and the superposition of vertices not adjacent to the marked vertex. \emph{Strongly regular graphs} have exactly the structure to support such evolution: one with parameters ($N$, $k$, $\lambda$, $\mu$) has $N$ vertices, each with $k$ neighbors, where adjacent vertices have $\lambda$ common neighbors and non-adjacent vertices have $\mu$ common neighbors.  This means that relative to a marked vertex, colored red in Fig.~\ref{fig:graphs}, there are $k$ adjacent vertices, colored blue, and $N-k-1$ vertices, all at distance 2, colored white.

As one might expect, for some parameters ($N$, $k$, $\lambda$, $\mu$), there are no strongly regular graphs. One necessary, but insufficient, constraint is that the parameters satisfy \cite{Cameron1991}
\begin{equation}
	k(k - \lambda - 1) = (N - k - 1)\mu,
	\label{eq:params}
\end{equation}
which is proved by counting the pairs of adjacent blue and white vertices. On the left hand side of (\ref{eq:params}), the marked red vertex has $k$ neighbors, so there are $k$ blue vertices. Each blue vertex has $k$ neighbors, one of which is the red marked vertex, and $\lambda$ of which are other blue vertices. So it is adjacent to $k - \lambda - 1$ white vertices. Thus the number of pairs of adjacent blue and white vertices is $k(k - \lambda - 1)$. On the right hand side of (\ref{eq:params}), we count the number of pairs another way, beginning with the white vertices. There are $N$ total vertices in the graph, one of which is red and $k$ of which are blue. So there are $N - k - 1$ white vertices. Each of these white vertices is adjacent to $\mu$ blue vertices. So there are $(N - k - 1)\mu$ pairs of blue and white vertices. Equating these expressions gives (\ref{eq:params}).

Equation \ref{eq:params} also implies that that $k$, the degree of the vertices, must be lower bounded by $\sqrt{N}$. That is, $k^2 > k(k-\lambda-1) = (N-k-1)\mu$, so
\begin{equation}
	k = \Omega(\sqrt{N}).
	\label{eq:kN}
\end{equation}

While not all strongly regular graphs are known, certain parameter families are. One family is the Paley graphs, which are parameterized by
\[ N = 4t + 1, \quad k = 2t, \quad \lambda = t - 1, \quad \text{and} \quad \mu = t, \]
where $N$ must be a prime power, and be congruent to $1\text{ mod }4$ \cite{Cameron1991}. The $t = 2$ case is shown in Fig.~\ref{fig:graphs}. Another family is the Latin square graphs, which are parameterized by
\[ N = t^2, \, k = d(t-1), \, \lambda = d^2 - 3d + t, \, \text{and} \, \mu = d(d-1), \]
with the additional condition that the graph be geometric, in the sense of finite geometries \cite{Cameron1991}. When $d = 3$, they can be pictured as a square lattice of $t^2$ vertices, where each vertex is given a symbol that only appears once in each row and column \cite{Babai1995}. An example of this is shown in Fig.~\ref{fig:graphs}. Vertices are connected if they are in the same row or column or have the same symbol.

Although it is not apparent from the small, symmetrical example(s) in Fig.~\ref{fig:graphs}, Latin square graphs are proved to be asymmetric for large $N$, meaning their automorphism groups are trivial \cite{Babai1995}, as are ``almost all'' strongly regular graphs in general, although a general proof seems unlikely \cite{Babai1995}. Thus they are not homogeneous (vertex transitive); there is no automorphism taking a vertex to any other vertex as there is for the complete graph, the hypercube, and cubical lattices, and which therefore might seem necessary for quantum random walk search to succeed \cite{Grover1996, CG2004}. We show this intuition to be false; a randomly walking quantum particle on strongly regular graphs optimally \cite{Zalka1999} solves the quantum search problem in $O(\sqrt{N})$ time for large $N$.

%-------------------------------------------------------------------------------
% Section
%-------------------------------------------------------------------------------

\section{Setup}

We begin by grouping the three types of vertices together: the red marked vertex, $k$ blue vertices that are adjacent to the red marked vertex, and $N - k - 1$ white vertices that are not adjacent to the red marked vertex. Call the respective equal superpositions of them $\ket{w}$, $\ket{a}$, and $\ket{b}$; they form a three-dimensional subspace of $\mathbb{C}^N$:
\[ \ket{w} = \begin{pmatrix} 1 \\ 0 \\ 0 \end{pmatrix}, \quad \ket{a} = \frac{1}{\sqrt{k}} \sum_{(i,w) \in \mathcal{E}} \ket{x} = \begin{pmatrix} 0 \\ 1 \\ 0 \end{pmatrix}, \]
\[ \ket{b} = \frac{1}{\sqrt{N-k-1}} \sum_{(i,w) \not\in \mathcal{E}} \ket{x} = \begin{pmatrix} 0 \\ 0 \\ 1 \end{pmatrix}. \]
The system begins in the equal superposition of all vertices $\ket{s}$, which we can write in the $\{\ket{w}, \ket{a}, \ket{b}\}$ basis:
\[ \ket{s} = \frac{1}{\sqrt{N}} \sum_x \ket{x} = \frac{1}{\sqrt{N}} \begin{pmatrix} 1 \\ \sqrt{k} \\ \sqrt{N-k-1} \end{pmatrix}. \]
The system evolves by Schr\"odinger's equation with the search Hamiltonian from (\ref{eq:QRW}). In the case of strongly regular graphs, each vertex has degree $k$, so the degree matrix is a multiple of the identity matrix: $D = kI$. This is simply a rescaling of energy, so we can drop it without observable effects. Then the Hamiltonian is $H = -\gamma A - \ketbra{w}{w}$. The $\ketbra{w}{w}$ term is simply a $3 \times 3$ matrix with a $1$ in the top-left corner and $0$'s everywhere else. The adjacency matrix $A$ is
\[ A = \begin{pmatrix}
	0 & \sqrt{k} & 0 \\
	\sqrt{k} & \lambda & \sqrt{\mu}\sqrt{k-\lambda-1} \\
	0 & \sqrt{\mu}\sqrt{k-\lambda-1} & k-\mu \\
\end{pmatrix}, \]
where the last item in the second row, for example, is $\sqrt{k} / \sqrt{N-k-1}$ to convert between the normalization of $\ket{b}$ and $\ket{c}$, times the $k-\lambda-1$ white vertices that go into a blue vertex (see Fig.~\ref{fig:graphs}), followed by simplification using (\ref{eq:params}). Thus the Hamiltonian is
\begin{equation}
	H = -\gamma \begin{pmatrix}
		\frac{1}{\gamma} & \sqrt{k} & 0 \\
		\sqrt{k} & \lambda & \sqrt{\mu}\sqrt{k-\lambda-1} \\
		0 & \sqrt{\mu}\sqrt{k-\lambda-1} & k-\mu \\
	\end{pmatrix}.
	\label{eq:Hamiltonian}
\end{equation}

\section{Solution using Perturbation Theory}

For the complete graph, the perturbation $H^{(1)}$ caused the eigenstates to be a linear combination of $\ket{w}$ and $\ket{r}$. To make this more clear for strongly regular graphs, we transform from the $\{\ket{w}, \ket{a}, \ket{b}\}$ basis to the $\{\ket{w}, \ket{r}, \ket{e_3}\}$ basis, where
\[ \ket{e_3} = \frac{1}{\sqrt{N-1}} \left( \sqrt{N-k-1} \ket{a} - \sqrt{k} \ket{b} \right). \]
We do this by conjugating (\ref{eq:Hamiltonian}) by
\[ T = \begin{pmatrix} \ket{w} & \ket{r} & \ket{e_3} \end{pmatrix} = \begin{pmatrix}
	1 & 0 & 0 \\
	0 & \frac{\sqrt{k}}{\sqrt{N-1}} & \frac{\sqrt{N-k-1}}{\sqrt{N-1}} \\
	0 & \frac{\sqrt{N-k-1}}{\sqrt{N-1}} & -\frac{\sqrt{k}}{\sqrt{N-1}}
\end{pmatrix}. \]
Multiplying $T^{-1} H T$, the Hamiltonian in the $\{\ket{w}, \ket{r}, \ket{e_3}\}$ basis is
\begin{equation}
	H = - \gamma \begin{pmatrix}
		\frac{1}{\gamma} & \frac{k}{\sqrt{N-1}} & \frac{\sqrt{k}\sqrt{N-k-1}}{\sqrt{N-1}} \\
		\frac{k}{\sqrt{N-1}} & \frac{k(N-2)}{N-1} & \frac{-\sqrt{k}\sqrt{N-k-1}}{N-1} \\
		\frac{\sqrt{k}\sqrt{N-k-1}}{\sqrt{N-1}} & \frac{-\sqrt{k}\sqrt{N-k-1}}{N-1} & \frac{(\lambda-\mu)(N-1) + k}{N-1} \\
	\end{pmatrix}.
	\label{eq:HamiltonianPrime}
\end{equation}
Now we break the problem into two cases: when $k$ scales as $N$ and when $k$ scales less than $N$---but still no less than $\sqrt{N}$ from (\ref{eq:kN}).

\begin{figure}
\begin{center}
	\includegraphics[height=1.2in]{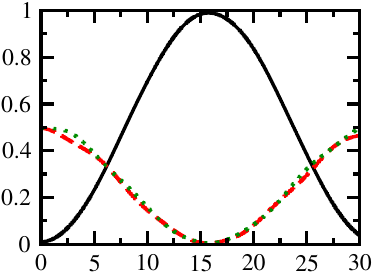}
	\includegraphics[height=1.2in]{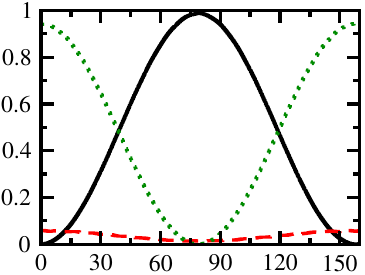}
	\caption{\label{fig:prob_time}Search on the Paley graph with parameters (101,50,24,25) with $\gamma_{c1}$ in (\ref{eq:gammac1}) (left), and Latin square graph with parameters (2500,147,50,6) with $\gamma_{c2}$ in (\ref{eq:gammac2}) (right). The black solid curve is $\left| \braket{w}{\psi} \right|^2$, the red dashed curve is $\left| \braket{a}{\psi} \right|^2$, and the green dotted curve is $\left| \braket{b}{\psi} \right|^2$.}
\end{center}
\end{figure}

\emph{Case 1: $k = \Theta(N)$.} The leading and first order terms of the Hamiltonian in (\ref{eq:HamiltonianPrime}) are, for large $N$,
\[ H^{(0)} = -\gamma \begin{pmatrix}
	\frac{1}{\gamma} & 0 & 0 \\
	0 & k & 0 \\
	0 & 0 & \lambda - \mu \\
\end{pmatrix}\!,
H^{(1)} = -\gamma \begin{pmatrix}
	0 & \frac{k}{\sqrt{N}} & \sqrt{k} \\
	\frac{k}{\sqrt{N}} & 0 & 0 \\
	\sqrt{k} & 0 & 0 \\
\end{pmatrix}\!. \]
Clearly, the eigenvectors of $H^{(0)}$ are $\ket{w}$, $\ket{r}$, and $\ket{e_3}$ with corresponding eigenvalues $-1$, $-\gamma k$, and $-\gamma(\lambda - \mu)$. We want $\ket{w}$ and $\ket{r}$ to be degenerate, which occurs when
\begin{equation}
	\gamma_{c1} = \frac{1}{k}.
	\label{eq:gammac1}
\end{equation}
Then the corresponding eigenstates of the perturbed system are $\ket{\psi_\pm} = \alpha_w \ket{w} + \alpha_r \ket{r}$. The coefficients $\alpha_w$ and $\alpha_r$ can be found by solving the eigenvalue problem
\[ \begin{pmatrix} H_{ww} & H_{wr} \\ H_{rw} & H_{rr} \end{pmatrix} \begin{pmatrix} \alpha_w \\ \alpha_r \end{pmatrix} = E_\pm \begin{pmatrix} \alpha_w \\ \alpha_r \end{pmatrix}, \]
where $H_{wr} = \langle w | H^{(0)} + H^{(1)} | r \rangle$. Evaluating the matrix components with $\gamma = \gamma_{c1} = 1/k$ and large $N$, we get
\[ \begin{pmatrix} -1 & \frac{-1}{N} \\ \frac{-1}{N} & -1 \end{pmatrix} \begin{pmatrix} \alpha_w \\ \alpha_r \end{pmatrix} = E_\pm \begin{pmatrix} \alpha_w \\ \alpha_r \end{pmatrix}. \]
Solving this, we get eigenstates $\ket{\psi_\pm} = \left( \ket{r} \mp \ket{w} \right)/\sqrt{2}$ with eigenvalues $E_\pm = -1 \pm 1/N$. Since $\ket{r} \approx \ket{s}$, the system evolves from $\ket{s}$ to nearly $\ket{w}$ in time $t_* = \pi / \Delta E = \pi \sqrt{N} / 2$ for large $N$. This is shown in Fig.~\ref{fig:prob_time}.

\textit{Case 2: $k = o(N)$.} The leading and first order terms of the Hamiltonian in (\ref{eq:HamiltonianPrime}) are, for large $N$,
\[ H^{(0)} = -\gamma \! \begin{pmatrix}
	\frac{1}{\gamma} & 0 & \sqrt{k} \\
	0 & k & 0 \\
	\sqrt{k} & 0 & \lambda-\mu \\
\end{pmatrix}\!, \,
H^{(1)} = \frac{-\gamma k}{\sqrt{N}} \! \begin{pmatrix}
	0 & 1 & 0 \\
	1 & 0 & 0 \\
	0 & 0 & 0 \\
\end{pmatrix}\!. \]
It's clear that $\ket{r}$ is an eigenvector of $H^{(0)}$ with eigenvalue $-\gamma k$. Then the two other eigenvectors have the form $\begin{pmatrix} c_1 & 0 & c_3 \end{pmatrix}^\intercal$. We want one of these to have the same eigenvalue $-\gamma k$ so that $H^{(0)}$ is degenerate:
\[ H^{(0)} \begin{pmatrix} c_1 \\ 0 \\ c_3 \end{pmatrix} = -\gamma k \begin{pmatrix} c_1 \\ 0 \\ c_3 \end{pmatrix}. \]
Solving this gives the critical $\gamma$ when $k = \Theta(\sqrt{N})$,
\begin{equation}
	\gamma_{c2} = \frac{1}{k} + \frac{1}{(N-1)\mu},
	\label{eq:gammac2}
\end{equation}
and corresponding eigenvector
\[ \ket{c} = \underbrace{\left( 1 + \frac{(k - \lambda + \mu)^2}{k} \right)^{-1/2}}_{C} \begin{pmatrix} \frac{k - \lambda + \mu}{\sqrt{k}} \\ 0 \\ 1 \end{pmatrix}, \]
where we've called the normalization constant $C$. Note
\begin{equation}
	k - \lambda + \mu = (N-k-1)\frac{\mu}{k} + \mu + 1 \approx \frac{\mu N}{k}, \label{eq:klm}
\end{equation}
using (\ref{eq:params}), $k = o(N)$, and large $N$. Then $C$ becomes
\begin{equation}
	C \approx \left( 1 + \frac{(\mu N)^2}{k^3} \right)^{-1/2} \approx \frac{k^{3/2}}{\mu N},
	\label{eq:C}
\end{equation}
when $k$ scales less than or equal to $(\mu N)^{2/3}$, which is true for the known parameter families of Latin square graphs, pseudo-Latin square graphs, negative Latin square graphs, square lattice graphs, triangular graphs, and point graphs of partial geometries \cite{Cameron1991}. $C$ is dominated by $1$ otherwise, for which we are unaware of any examples (when $k$ scales less than $N$).

The perturbation causes the eigenstates of $H^{(0)} + H^{(1)}$ to be a linear combination of $\ket{r}$ and $\ket{c}$: $\ket{\psi_\pm} = \alpha_r \ket{r} + \alpha_c \ket{c}$. To find $\alpha_r$ and $\alpha_c$, we solve the eigenvalue problem
\[ \begin{pmatrix} H_{rr} & H_{rc} \\ H_{cr} & H_{cc} \end{pmatrix} \begin{pmatrix} \alpha_w \\ \alpha_r \end{pmatrix} = E_\pm \begin{pmatrix} \alpha_w \\ \alpha_r \end{pmatrix}, \]
where $H_{rc} = \langle r | H^{(0)} + H^{(1)} | c \rangle$, \textit{etc}. These terms are straightforward to calculate. We get
\[ \begin{pmatrix} -\gamma k & -\gamma C \sqrt{\frac{N}{k}} \mu \\ -\gamma C \sqrt{\frac{N}{k}} \mu & -\gamma k \end{pmatrix} \begin{pmatrix} \alpha_w \\ \alpha_r \end{pmatrix} = E_\pm \begin{pmatrix} \alpha_w \\ \alpha_r \end{pmatrix}, \]
where for the off-diagonal terms we used (\ref{eq:params}) and $(N-1) \approx N$.
Solving this, the eigenstates of $H' = H^{(0)} + H^{(1)}$ are $\ket{\psi_\pm} = \left( \ket{r} \mp \ket{c} \right)/\sqrt{2}$ with eigenvalues $E_\pm = -\gamma k \pm \gamma C \mu \sqrt{N/k}$.

Now let's find the success probability as a function of time. Solving Schr\"odinger's equation, the evolution of the system is approximately $\ket{\psi(t)} \approx e^{-iH't} \ket{s}$. The state of the system approximately evolves in the subspace spanned by $\ket{\psi_\pm}$ for large $N$, so this becomes
\[ \ket{\psi(t)} \approx e^{-iE_+t} \ket{\psi_+} \braket{\psi_+}{s} + e^{-iE_-t} \ket{\psi_-} \braket{\psi_-}{s}. \]
Note that $\braket{\psi_\pm}{s} = \left( \braket{r}{s} \mp \braket{c}{s} \right)/\sqrt{2} \approx (1 \mp 0)/\sqrt{2} = 1/\sqrt{2}$ for large $N$. Multiplying by $\bra{w}$ on the left, noting $\braket{w}{\psi_\pm} = \mp \braket{w}{c}/\sqrt{2} \approx \mp \frac{1}{2} C \mu N / k^{3/2}$ from (\ref{eq:klm}), and plugging in the energy eigenvalues, the success amplitude is
\[ \braket{w}{\psi(t)} \approx e^{-i \gamma k t} \frac{1}{2} C \frac{\mu N}{k^{3/2}} \left( -e^{-i \gamma C \sqrt{N/k} \mu t} + e^{i \gamma C \sqrt{N/k} \mu t} \right). \]
The exponentials sum to $2i\sin(\cdot)$, so the success probability is
\[ \left| \braket{w}{\psi(t)} \right|^2 \approx \left( C \frac{\mu N}{k^{3/2}} \right)^2 \sin^2 \left( C \frac{\mu\sqrt{N}}{k^{3/2}} t \right), \]
where we've used $\gamma \approx 1/k$. For the known parameter families where $k = O((\mu N)^{2/3})$, which includes Latin square graphs that are proved asymmetric \cite{Babai1995}, we use (\ref{eq:C}) to get $\left| \braket{w}{\psi(t)} \right|^2 \approx \sin^2 ( t/\sqrt{N} )$, so the search is achieved with probability $1$ in time $t_* = \pi \sqrt{N} / 2$ for large $N$, as shown in Fig.~\ref{fig:prob_time}.

Thus we've shown that quantum search on known strongly regular graphs behaves like search on the complete graph for large $N$, reaching a success probability of $1$ at time $\Theta(\sqrt{N})$. This requires choosing $\gamma = \gamma_{c1} = 1/k$ when $k = \Theta(N)$ and $\gamma_{c2} = 1/k + 1/[(N-1)\mu]$ when $k = o(N)$. Since this includes strongly regular graphs that are asymmetric, it disproves the intuition that fast quantum search requires global symmetry.

\begin{acknowledgments}
	This work was partially supported by the Defense Advanced Research Projects Agency as part of the Quantum Entanglement Science and Technology program under grant N66001-09-1-2025, the Air Force Office of Scientific Research as part of the Transformational Computing in Aerospace Science and Engineering Initiative under grant FA9550-12-1-0046, and the Achievement Awards for College Scientists Foundation. We thank Joy Morris for useful discussions.
\end{acknowledgments}

%-------------------------------------------------------------------------------
% References.
%-------------------------------------------------------------------------------

\bibliography{refs}

\end{document}